# Magnetization switching induced by spin-orbit torque from Co$_2$MnGa magnetic Weyl semimetal thin films

Ke Tang,[1,2] Zhenchao Wen,[1,*] Yong-Chang Lau,[3,4] Hiroaki Sukegawa,[1] Takeshi Seki,[3,4] and Seiji Mitani[1,2]

[1]*National Institute for Materials Science (NIMS), Tsukuba 305-0047, Japan*
[2]*Graduate School of Pure and Applied Sciences, University of Tsukuba, Tsukuba 305-8577, Japan*
[3]*Institute for Materials Research, Tohoku University, Sendai 980-8577, Japan*
[4]*Center for Spintronics Research Network, Tohoku University, Sendai 980-8577, Japan*

**Abstract**

This study reports the magnetization switching induced by spin-orbit torque (SOT) from the spin current generated in Co$_2$MnGa magnetic Weyl semimetal (WSM) thin films. We deposited epitaxial Co$_2$MnGa thin films with highly *B*2-ordered structure on MgO(001) substrates. The SOT was characterized by harmonic Hall measurements in a Co$_2$MnGa/Ti/CoFeB heterostructure and a relatively large spin Hall efficiency ($\xi_{SH}$) of −7.8% was obtained. The SOT-induced magnetization switching of the perpendicularly magnetized CoFeB layer was further demonstrated using the structure. The symmetry of second harmonic signals, thickness dependence of $\xi_{SH}$, and shift of anomalous Hall loops under applied currents were also investigated. This study not only contributes to the understanding of the mechanisms of spin-current generation from magnetic-WSM-based heterostructures, but also paves a way for the applications of magnetic WSMs in spintronic devices.

*Corresponding author: WEN.Zhenchao@nims.go.jp

Magnetization switching induced by spin-orbit torques (SOT) is a key technology for achieving ultrafast and energy-efficient memory devices, i.e. magnetoresistive random-access memories.[1] SOTs generated from ferromagnetic materials (FMs) have drawn significant attention because of their remarkable charge-to-spin conversion efficiency and the possibility of manipulating the direction of the spin polarization carried by the spin current.[2–14] Studies have reported the inclusion of bulk and interface contributions in the mechanisms of spin-current generation from FMs.[15,16] The spin anomalous Hall effect (SAHE), which is related to the anomalous Hall effect (AHE), is one of the bulk contributions.[4,15] The bulk-generated spin current via SAHE is polarized with magnetization and flows in the **m** × **E** direction,[15] where **m** and **E** are the FM's moment vector and electric field vector, respectively. The **m**-independent SHE is another bulk contribution,[17] which has the same symmetry as the observed SHE in nonmagnetic materials. In contrast, the interface-generated spin current originates from the spin-orbit-dependent scattering processes at the interface, i.e., spin-orbit filtering and spin-orbit precession, thus resulting in spin currents with the spin polarization along the **z** × **E** and **m** × **y** directions, respectively,[8,16,18] where **E** is applied along the **x**-direction. Recently, in the CoFeB/Ti/CoFeB trilayers, magnetization switching induced by SOTs was achieved, where the generated spin current at the interface of the bottom layer, CoFeB/Ti, traverses the Ti layer and exerts a torque at the top-CoFeB layer.[8] More recently, a large SAHE was observed in the $L1_0$-FePt films, and SAHE-induced magnetization switching was demonstrated in a giant magnetoresistance device having a FePt/Cu/NiFe trilayer structure.[12] These results indicated that SOTs from FMs are promising for applications and exotic FMs exhibiting large and useful spin-orbit effects are strongly required.

Magnetic Weyl semimetal (WSM) is a family of topological materials with three dimensional linearly dispersive band crossing points, i.e. Weyl nodes, owing to time-reversal-symmetry breaking.[19]

Recently, large intrinsic spin-Hall conductivities in WSMs were predicted by the first-principles calculations.[20] Because semimetals usually have low charge conductivities, the WSMs may show a larger spin-Hall efficiency compared to heavy metals. Moreover, because of the existence of spontaneous magnetization in magnetic WSMs, the manipulation of the direction of spin polarization for spin current becomes possible and a field-free SOT switching could be realized by using the magnetic WSMs. In addition, the enhancement of Gilbert damping usually happens in the heavy metal/FM structures due to the spin absorption by heavy metal.[21] The damping enhancement could be smaller in WSM-based structures because of the absence of heavy elements. Recently, topological Weyl states in magnetic $Co_2MnGa$ and $Co_3Sn_2S_2$ compounds were visualized using photoemission experiments.[22,23] Large AHEs were observed in $Co_2MnGa$ and $Co_3Sn_2S_2$ because of enhanced Berry curvatures near the Weyl nodes.[24–27] To date, the spin-current generation and SOT-induced magnetization switching, however, have not been reported experimentally in magnetic WSMs. In this work, we have fabricated heterostructures based on ferromagnetic $Co_2MnGa$ WSM films and performed harmonic Hall measurement to study the spin currents and SOTs generated from the $Co_2MnGa$ thin films. The SOT-induced magnetization switching is achieved in SOT devices whose core structure consisted of $Co_2MnGa$/Ti/CoFeB trilayers.

The $Co_2MnGa$ films were deposited on MgO(001) single-crystalline substrates by magnetron sputtering under a base pressure of approximately $2 \times 10^{-6}$ Pa. The deposition temperature ($T_{depo}$) varied from 270 to 560 °C. The composition of the films was estimated to be $Co_2Mn_{1.14}Ga_{0.95}$ by X-ray fluorescence spectroscopy. The surface structure and morphology of the films were evaluated by *in-situ* reflection high energy electron diffraction (RHEED) and *ex-situ* atomic force microscope (AFM), respectively. Out-of-plane and in-plane X-ray diffraction (XRD) were used to characterize the crystal structure. Magnetic properties were measured using a vibrating sample magnetometer (VSM). A heterostructure consisting of $Co_2MnGa$/Ti/CoFeB/MgO was fabricated in order to evaluate the spin-transport properties. The heterostructure was post-annealed at 300 °C for 30 min and was patterned into Hall bar structures using UV lithography and Ar ion milling. Ta(5)/Au(150) (layer thicknesses are in nanometers) layers were then deposited on the Hall bars as electrodes by lift-off process. The harmonic Hall measurement and SOT-induced magnetization switching were then performed in a physical properties measurement system (PPMS) at room temperature.

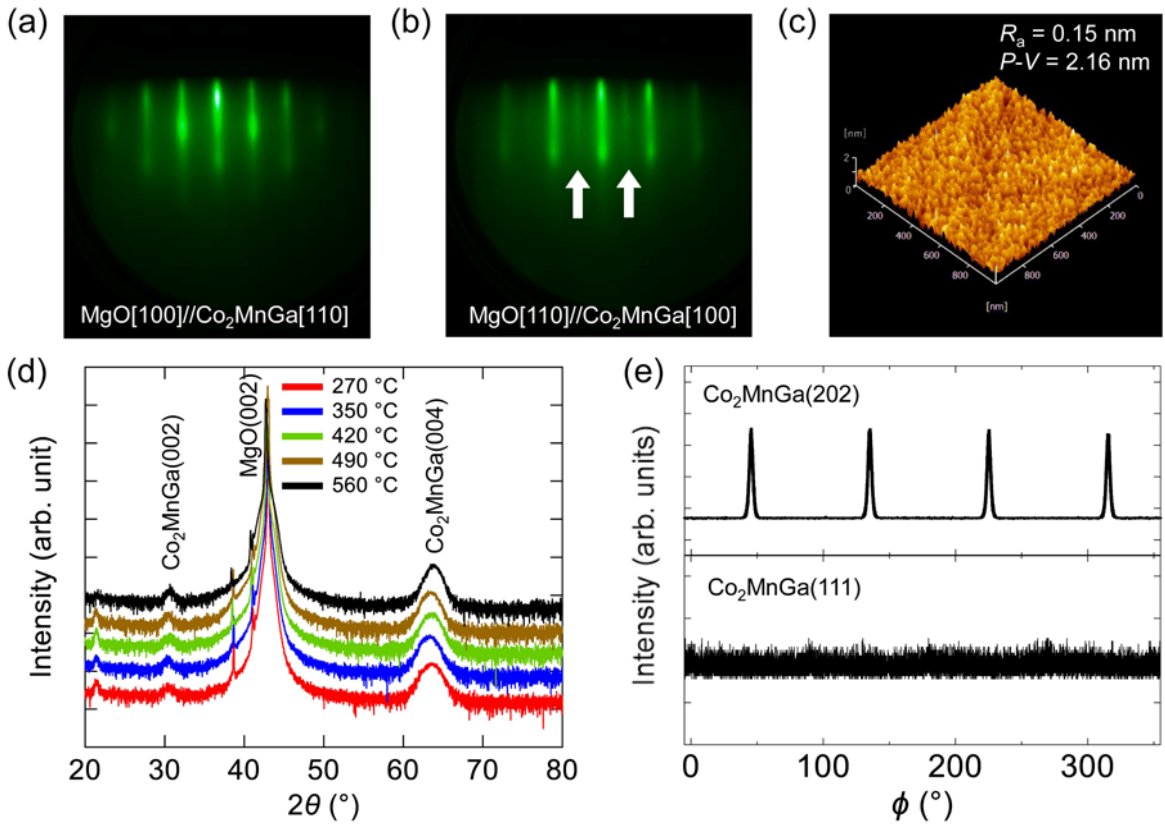

**FIG. 1.** (a) and (b) RHEED patterns for the surface of a 10-nm-thick $Co_2MnGa$ film with the electron beam parallel to MgO[100] and MgO[110] azimuths, respectively. The white arrows indicate the superlattice streaks. (c) AFM image for the $Co_2MnGa$ (10 nm)/$Mg_2AlO_x$ (2 nm) surface. (d) Out-of-plane XRD patterns of the thin films deposited at varied $T_{depo}$ between 270 °C and 560 °C. (e) In-plane ($\phi$-scan) patterns of the $Co_2MnGa$(202) and $Co_2MnGa$(111) planes. The $Co_2MnGa$ film characterized in (a), (b), (c), and (e) was deposited at $T_{depo}$ = 560 °C.

Figures 1(a) and 1(b) show the RHEED patterns taken at the surface of a 10-nm-thick $Co_2MnGa$ film deposited at 560 °C. The electron beam incident direction was parallel to the MgO[100] and MgO[110] azimuths, corresponding to the $Co_2MnGa$[110] and $Co_2MnGa$[100], respectively. The sharp and narrow streaks of both patterns indicate the epitaxial growth of the $Co_2MnGa$(001) film. In Fig. 1(b), the white arrows denote the $B2$ superlattice streaks. Figure 1(c) shows the AFM images of the $Co_2MnGa$ film capped by an $Mg_2AlO_x$ (2 nm) layer. The average roughness of the AFM image was obtained to be 0.15 nm, indicating a flat film surface that is suitable as an underlayer for the Ti/CoFeB stacks. Figure 1(d) shows the out-of-plane XRD patterns of 10-nm-thick $Co_2MnGa$ films deposited between 270 °C and 560 °C. Besides the peaks from MgO(001) substrates, the peaks of the $Co_2MnGa$(002) and $Co_2MnGa$(004) lattice planes were observed in the out-of-plane patterns, indicating that $Co_2MnGa$ films were grown with the (001) orientation. In Fig. 1(e), the pole scans were carried out by titling the films angle $\chi$ at approximately 45° for (202) (upper panel) and approximately 54.7° for (111) lattice planes (lower panel). The $Co_2MnGa$(202) peaks show a four-fold rotational symmetry, confirming the epitaxial growth with (001) orientation of the $Co_2MnGa$ thin film. In addition, the absence of peaks from the $Co_2MnGa$(111) plane indicates that the $Co_2MnGa$ film has neither $L2_1$ nor $D0_3$ structure but has a $B2$ structure, where Mn and Ga sites are randomly occupied. The degree of ordering parameter for the $B2$ structure ($S_{B2}$) can be evaluated from the integrated intensities of the XRD peaks.[26] We obtained $S_{B2} \sim 97\%$ by assuming stoichiometric $Co_2MnGa$ composition for the film deposited at $T_{depo}$ = 560 °C.

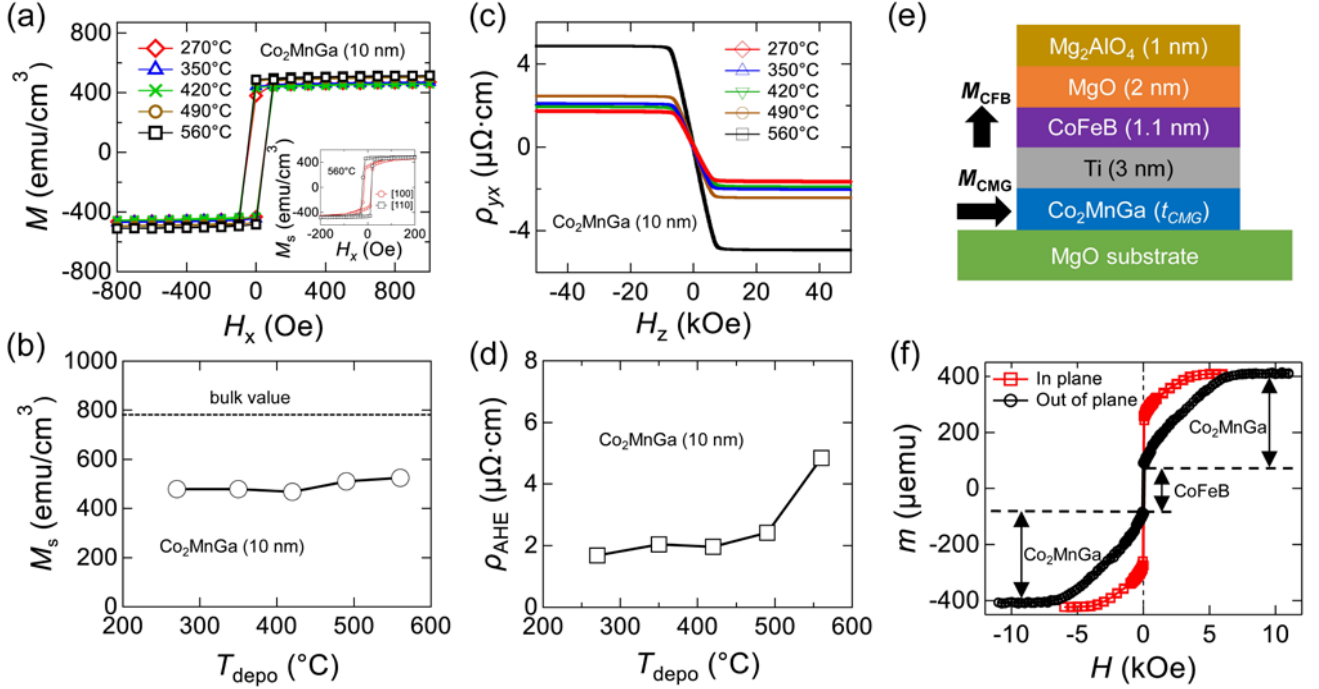

**FIG. 2.** (a) In-plane magnetic hysteresis loops of MgO substrate//Co$_2$MnGa (10 nm)/Mg$_2$AlO$_x$ (2 nm) along Co$_2$MnGa[110], where the Co$_2$MnGa layers were deposited at various $T_{\text{depo}}$. Inset shows the comparison of magnetic hysteresis loops with the magnetic field along Co$_2$MnGa[100] and [110]. (b) $T_{\text{depo}}$ dependence of the saturation magnetization $M_s$ of Co$_2$MnGa films. (c) Out-of-plane magnetic field ($H_z$) dependence of the transverse resistivity ($\rho_{yx}$). (d) $T_{\text{depo}}$ dependence of anomalous Hall resistivity of Co$_2$MnGa films. (e) The multilayer structure of the SOT devices used in the spin-transport measurements, where the Co$_2$MnGa layer was deposited at 560 ℃ with thickness $t_{\text{CMG}}$ ranging from 0 to 12.2 nm. (f) Out-of-plane and in-plane magnetic hysteresis loops of Co$_2$MnGa (6.1)/Ti (3)/CoFeB (1.1)/MgO (2)/Mg$_2$AlO$_x$ (1) (unit in nm) heterostructure. The arrows denote the contributions of Co$_2$MnGa and CoFeB to the magnetic moments in the out-of-plane measurement.

Figure 2(a) shows the in-plane magnetic field dependence of the magnetization for the Co$_2$MnGa films deposited at different temperatures evaluated by VSM at room temperature, where the magnetic field is along Co$_2$MnGa[110]. Typical ferromagnetic hysteresis loops were observed. The comparison of magnetic hysteresis loops with the magnetic field along Co$_2$MnGa[100] and [110] is presented in the inset of Fig. 2(a), which indicates the magnetic easy axis of the Co$_2$MnGa film is along the [110] direction. The saturation magnetization ($M_s$) slightly increases with $T_{\text{depo}}$, as shown in Fig.

2(b). The maximum $M_s$ is obtained to be ~500 emu/cm$^3$ for the Co$_2$MnGa films, which is smaller than the reported value ($M_s$ ~ 780 emu/cm$^3$) of the bulk Co$_2$MnGa samples.[24] This may be the result of degraded chemical ordering and off-stoichiometric composition. The AHE of the Co$_2$MnGa films was further characterized by measuring Hall resistivity under an out-of-plane magnetic field $H_z$, as shown in Fig. 2(c). It is found that the sign of AHE is positive (hole like) in the Co$_2$MnGa, which is consistent with the previous report.[24,26] In Fig. 2(d), the derived anomalous Hall resistivity ($\rho_{AHE}$) is plotted as a function of $T_{depo}$; $\rho_{AHE}$ increases with the increase of $T_{depo}$ and shows the maximum value of 4.8 μΩ cm at $T_{depo}$ = 560 °C. The $\rho_{AHE}$ of nearly stoichiometric Co$_2$MnGa films with an $L2_1$-ordered structure was reported to be approximately 17 μΩ cm.[26] In the present Co$_2$MnGa films, the absence of $L2_1$ order may lead to reduced $\rho_{AHE}$. Figure 2(e) schematically shows the heterostructure consisting of Co$_2$MnGa ($t_{CMG}$)/Ti (3)/CoFeB (1.1)/MgO (2)/Mg$_2$AlO$_x$ (1) (unit: nm) for investigating spin transport properties and SOT-induced magnetization switching, where the Co$_2$MnGa layer was deposited at 560 °C. The thickness of the Co$_2$MnGa layer ($t_{CMG}$) ranges from 0 to 12.2 nm. The CoFeB layer is perpendicularly magnetized and serves as a spin-current detector. The Ti layer is inserted to prevent the direct coupling between Co$_2$MnGa and CoFeB. In addition, the spin Hall efficiency of Ti was reported to be negligible, which would not affect the evaluation of the spin transport behavior of the bottom FM layer.[8] Figure 2(f) indicates the magnetic hysteresis loops of the heterostructure with $t_{CMG}$ = 6.1 nm. The double-head arrows denote the moment contributions of the Co$_2$MnGa and CoFeB layers in the out-of-plane hysteresis loop. The Co$_2$MnGa layer shows a magnetic hard axis, whereas the CoFeB layer has a magnetic easy axis under the out-of-plane magnetic field. Thus, we achieved the perpendicularly magnetized CoFeB layer, which is necessary for detecting the SOT-induced magnetization switching by anomalous Hall measurement.

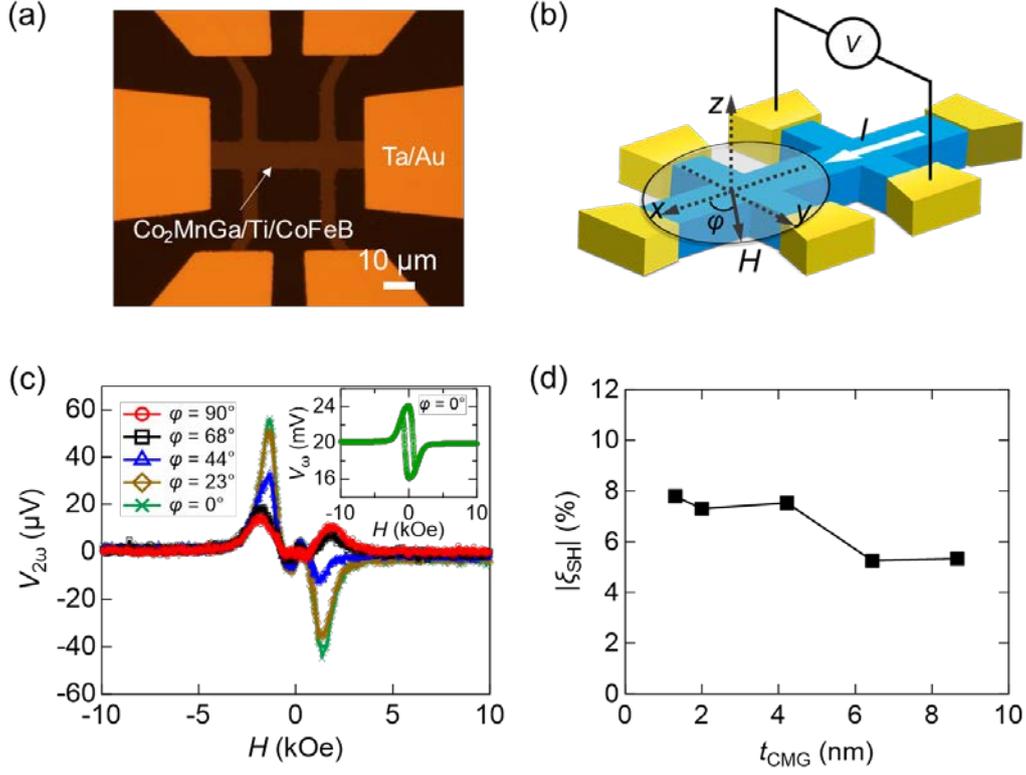

**FIG. 3.** (a) Photomicroscope image of a Hall bar device used in this study, where the long axis of the Hall bar is along $Co_2MnGa[110]$. (b) The coordination and configuration of harmonic Hall measurement. (c) The second harmonic Hall voltage signals at different azimuthal angles $\varphi$ for a sample with $Co_2MnGa$ (1.3 nm)/Ti (3 nm)/CoFeB (1.1 nm) layer structure. Inset is the first harmonic Hall signals at $\varphi = 0°$. (d) $t_{CMG}$ dependence of the absolute value of $\xi_{SH}$.

Multilayer stacks with the core structure of $Co_2MnGa$/Ti/CoFeB were microfabricated into the Hall bar structures, as shown in Fig. 3(a). The width and length of the Hall bar are 10 μm and 25 μm, respectively. The harmonic Hall measurement was performed in order to investigate the spin-orbit torques in the heterostructures. The coordination and configuration of the measurements are indicated in Fig. 3(b). A sinusoidal current ($I$) with a constant amplitude of 4 mA and a frequency of 512.32 Hz with an external magnetic field in the $xy$ plane. Figure 3(c) shows the second harmonic Hall voltage $V_{2\omega}$ against the magnetic field for different azimuthal angles, measured in a Hall bar device with $t_{CMG}$ = 1.3 nm. The effect of SOTs can be regarded as two effective fields, antidamping-like field ($H_{DL}$) and field-like field ($H_{FL}$), which can be given by the following equations[28]:

$$H_{DL} = -2\frac{H_{DL}^* \pm 2\epsilon H_{FL}^*}{1-4\epsilon^2}, \quad (1)$$

$$H_{FL} = -2\frac{H_{FL}^* \pm 2\epsilon H_{DL}^*}{1-4\epsilon^2}, \quad (2)$$

where "±" corresponds to the direction of magnetization $\pm M_z$, and $\epsilon$ represents the ratio of planar Hall resistance to anomalous Hall resistance, which we estimated to be 0.02. The $H_{DL}^*$ and $H_{FL}^*$ can be extracted from the harmonic Hall voltages in a low-field range by following equations:

$$H_{DL}^* \equiv \frac{\partial V_{2\omega}}{\partial H_x} / \frac{\partial^2 V_{\omega}}{\partial H_x^2}, \quad (3)$$

$$H_{FL}^* \equiv \frac{\partial V_{2\omega}}{\partial H_y} / \frac{\partial^2 V_{\omega}}{\partial H_y^2}, \quad (4)$$

where $V_{\omega}$ and $V_{2\omega}$ denote the first and second harmonic voltages, respectively, and $H_x$ and $H_y$ are the in-plane external magnetic fields along $x$ ($\varphi = 0°$) and $y$ ($\varphi = 90°$) axes. We obtained $H_{DL} = 7.2$ Oe and $H_{FL} = 1.4$ Oe from the harmonic Hall signals after subtracting the contributions of thermoelectric effect and Oersted field. The effective spin Hall efficiency ($\xi_{SH}$) is given by the following equation:

$$\xi_{SH} \equiv \frac{2e}{\hbar}\frac{H_{DL}M_s t_{CoFeB}}{j_0}, \quad (5)$$

where $e$ is the electron charge, $\hbar$ is the reduced Planck constant, $M_s$ is the saturated magnetization of CoFeB, $t_{CoFeB}$ is the thickness of the CoFeB layer, and $j_0$ is the current density in $Co_2MnGa$ after considering the resistivity of each layer. We evaluated the $\xi_{SH}$ to be −7.8% for the $Co_2MnGa$ ($t_{CMG} = 1.3$ nm)/Ti/CoFeB sample. A controlled sample with the structure of Ti (3 nm)/CoFeB (1.1 nm)/MgO (2 nm) was also fabricated for investigating the SHE of Ti layer. It was found that the spin Hall efficiency of Ti is approximately −0.3%, which is consistent with previous reports[8,29] and indicates the effect of Ti layer is negligible in the $Co_2MnGa$/Ti/CoFeB system. The thickness dependence of the absolute value of $\xi_{SH}$ is plotted in Fig. 3(d) where the amplitude of $\xi_{SH}$ decreases with the increase of $t_{CMG}$.

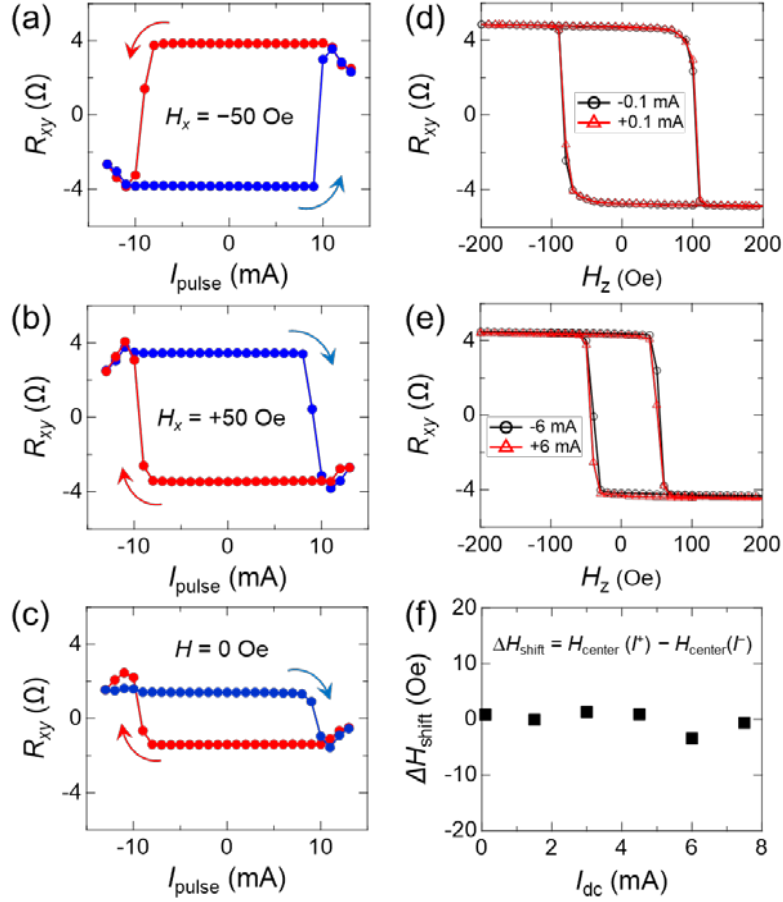

**FIG. 4.** (a) and (b) Current-induced magnetization switching with an external in-plane field $H_x = -50$ Oe and $H_x = +50$ Oe, respectively, for the $Co_2MnGa$ (1.3 nm)/Ti/CoFeB sample. (c) Current-induced magnetization switching without external magnetic field. The anomalous Hall resistance was measured using a DC current of 100 μA after applying a current pulse with 10-ms duration. (d) $R_{xy}$ versus $H_z$ curves with a current of ±0.1 mA. (e) $R_{xy}$ versus $H_z$ curves with a current of ±6 mA. (f) The shifts of hysteresis loops as a function of applied current.

The SOT-induced magnetization switching was demonstrated in the devices by applying a charge current in the Hall bar while monitoring the out-of-plane magnetization by Hall resistance. Figure 4(a) shows the magnetization switching with a small auxiliary magnetic field $H_x = -50$ Oe. The switching current is approximately 10 mA, corresponding to a current density of $J = 9.8 \times 10^6$ A/cm$^2$ in the $Co_2MnGa$ layer. By comparing the Hall resistance in the $R_{xy}$–$H_z$ loops, we found that the current reversed almost all the magnetic moments of CoFeB. When the magnetic field $H_x$ becomes opposite (+50 Oe), the switching behavior also reverses, as shown in Fig. 4(b). The demonstration of deterministic SOT switching with such a small $H_x$ bias field indicates that the Dzyaloshinskii–Moriya interaction (DMI) in Ti/CoFeB/MgO is relatively small. The magnetization of the CoFeB layer is preferred along the +z direction when the charge current flow and $H_x$ are parallel, suggesting that the sign of $\xi_{SH}$ is negative; that is, the spin polarization vector of the spin current is pointing along the +y direction at $Co_2MnGa$/Ti interface. We also investigated the field-free magnetization switching shown in Fig. 4(c). According to the value of $R_{xy}$, about half of the magnetic moments of the perpendicularly magnetized CoFeB is switched by the current, which may be due to the multi-domain structures in the CoFeB layer. In addition, the interlayer exchange coupling between $Co_2MnGa$ and CoFeB is not sufficient to compensate the DMI effective field which is probably small but may contribute to form multidomain structures. The DMI could be one of possible reasons for the partial SOT switching under zero field. To investigate the mechanisms of the SOT and spin current in the heterostructure, we measured the AHE with varying DC current amplitudes ($I_{DC}$) ranging from ±1.5 to ±7.5 mA, after the magnetic moment of $Co_2MnGa$ layer was aligned to the +**x** direction. We defined $H_{center}(I^+)$ and $H_{center}(I^-)$ as the centers of $R_{xy}$–$H_z$ loops at the positive and negative currents, respectively. The shift of the centers can be given by $\Delta H_{shift} = H_{center}(I^+) - H_{center}(I^-)$. Figures 4(d) and 4(e) show the $R_{xy}$–$H_z$ loops with $I_{DC} = \pm 0.1$ and ±6 mA, respectively. We observed that the switching field reduces with the increasing current and a very slight $\Delta H_{shift}$ at $I_{DC} = \pm 6$ mA. Figure 4(f) shows the current dependence of $\Delta H_{shift}$. The $\Delta H_{shift}$ is close to zero with the current in the measurement range. The small $\Delta H_{shift}$ corresponds to a tiny z component of the spin current, which may account for the partially field-free magnetization switching shown in Fig. 4(c).

We further discuss the origin of the spin-current generation and SOT switching in the Co$_2$MnGa-based heterostructures. Taniguchi *et al.* reported the bulk SAHE for spin-current generation in FMs, where the spin current with the spin polarization aligned along the **m** of the FM layer would flow in the **m** × **E** direction.[15] When **m** is parallel to **E**, the SAHE will not contribute to the spin-current generation. In our experiments, however, we observed non-zero second harmonic signal when $\varphi = 0°$ (equivalent to **m**∥**E**), as shown in Fig. 3(c), suggesting that the spin current in Co$_2$MnGa/Ti/CoFeB does not originate from the SAHE. Another bulk contribution is **m**-independent SHE,[17] which tends to increase with the increasing film thickness over a length scale related to the spin diffusion length, λ. However, we did not observe such an increase in the spin Hall efficiency, which may be due to the limited $t_{CMG}$ in our experiments. If **m**-independent SHE is responsible, this would imply that λ of Co$_2$MnGa is much lower than 1.3 nm. To elucidate the spin-current generation and SOTs in our Co$_2$MnGa/Ti/CoFeB structures, we also considered the interfacial effects. Amin *et al.* proposed the spin-current generation by the spin-orbit filtering and spin-orbit precession at the FM/nonmagnetic metal interface.[18] The former brings about the spins along **y** = **z** × **E** direction, whereas the latter gives spins along **z** = **m** × **y** direction. Baek *et al.* demonstrated the *z* component of the spin current by measuring the shift of AHE loops under positive and negative currents.[8] In our sample, the $\Delta H_{shift}$ is nearly negligible, indicating that the contribution of spin-orbit precession at the Co$_2$MnGa/Ti interface to spin-current generation is very small. Therefore, the spin-orbit filtering process at the interface with the resulting spins along **y** = **z** × **E** direction could be one of the origins for the spin-current generation and SOT-induced magnetization switching in the Co$_2$MnGa/Ti/CoFeB heterostructure. Furthermore, we consider intrinsic and scattering contributions to the spin-current generation. It was reported that the bulk topological band structure of Weyl semimetals could produce to a strong intrinsic spin-Hall effect.[20] The intrinsic mechanism may contribute to the **m**-independent SHE in our samples. In addition to the intrinsic mechanism, phonon skew scatterings[30] and/or magnon-excitation-related scatterings[31] may also contribute to the spin-current generation in **m**-independent SHE and interface effects because the SOT measurements were performed at room temperature. In order to reduce the current density of magnetization switching and achieve complete field-free magnetization switching, $Co_2MnGa$ films with stoichiometric composition and a highly $L2_1$-ordered structure may be required to introduce a large SAHE with controllable spin-polarization direction of the spin current.

In conclusion, flat and highly $B$2-ordered $Co_2MnGa$(001) magnetic WSM thin films were epitaxially grown by magnetron sputtering at varying growth temperatures. The spin-current generation and SOT-induced magnetization switching were achieved in Hall bar devices with $Co_2MnGa$(001)/Ti/CoFeB layer structure. A relatively large spin Hall efficiency up to −7.8% was observed in the well-controlled $Co_2MnGa$-based heterostructure by harmonic Hall measurements. The azimuthal angle dependence of second harmonic signals, the thickness dependence of spin Hall efficiency, and the shift of AHE loops with applied current demonstrated that the spin current for the SOT-induced magnetization switching might originate from spin-orbit filtering at the $Co_2MnGa$/Ti interface and/or **m**-independent SHE in $Co_2MnGa$. This study indicates that the $Co_2MnGa$ magnetic WSM thin film may be promising for SOT-based spintronic applications.

This work was partially supported by the KAKENHI (No. JP20K04569, No. JP20H00299, No. JP16H06332, and No. JP20K15156) from the Japan Society for the Promotion of Science (JSPS), the Inter-University Cooperative Research Program of the Institute for Materials Research, Tohoku University (No. 20K0058). K. Tang acknowledges National Institute for Materials Science for the provision of a NIMS Junior Research Assistantship.